\documentclass{iaus}
\usepackage{graphicx}

\title[progenitor of type Ia supernovae] 
{The single degenerate channel for the progenitor of type Ia
supernovae}

\author[Meng, Chen \& Han]   
{Xiangcun Meng$^{\rm 1, 2}$, Xuefei Chen$^{\rm 1}$ \& Zhanwen Han$^{\rm 1}$%
 }

\affiliation{$^1$National Astronomical Observatories/Yunnan
Observatory, the Chinese Academy of Sciences, Kunming, 650011,
China, \break email: conson859@msn.com\\[\affilskip]
$^2$Graduate School of the Chinese Academy of Sciences}

\pubyear{2008}
\volume{252}  
\pagerange{119--123}
\date{?? and in revised form ??}
\jname{Proceedings Title IAU Symposium}
\editors{A.C. Editor, B.D. Editor \& C.E. Editor, eds.}
\begin{document}

\maketitle

\begin{abstract}
We have carried out a detailed study of the single-degenerate
channel for the progenitors of type Ia supernovae (SNe Ia). In the
model, a carbon-oxygen white dwarf (CO WD) accretes hydrogen-rich
material from an unevolved or a slightly evolved non-degenerate
companion to increase its mass to Chandrasekhar mass limit.
Incorporating the prescription of \cite{HAC99a} for the accretion
efficiency into Eggleton's stellar evolution code and assuming
that the prescription is valid for all metallicities, we performed
binary stellar evolution calculations for more than 25,000 close
WD binary systems with various metallicities. The initial
parameter spaces for SNe Ia are presented in an orbital
period-secondary mass ($\log P_{\rm i}, M_{\rm 2}^{\rm i}$) plane
for each $Z$.

Adopting the results above, we studied the birth rate of SNe Ia
for various $Z$ via binary population synthesis. From the study,
we see that for a high $Z$, SNe Ia occur systemically earlier and
the peak value of the birth rate is larger if a single starburst
is assumed. The Galactic birth rate from the channel is lower than
(but comparable to) that inferred from observations.

We also showed the distributions of the parameters of the binary
systems at the moment of supernova explosion and the distributions
of the properties of companions after supernova explosion. The
former provides physics input to simulate the interaction between
supernova ejecta and its companion, and the latter is helpful to
search for the companions in supernova remnants.

\end{abstract}

\firstsection 
\section{Introduction}\label{sect:1}
Although type Ia supernovae (SNe Ia) appear to be good
cosmological distance indicators and have been applied
successfully in determining cosmological parameters (e.g. $\Omega$
and $\Lambda$; \cite{RIE98}; \cite{PER99}), the exact nature of
SNe Ia is still unclear, especially the progenitor model of SNe Ia
(see the reviews by \cite{HN00}; \cite{LEI00}). At present, the
single-degenerate Chandrasekhar model (\cite{WI73}; \cite{NTY84})
is the most widely accepted model. In the model, a CO WD accretes
hydrogen-rich material from its companion until its mass reaches a
mass of $\sim 1.378 M_{\odot}$ (close to Chandrasekhar mass,
\cite{NTY84}), and then explodes as a SN Ia. The companion is
probably a main sequence star or a slightly evolved star (WD+MS).
The discovery of the potential companion of Tycho's supernova also
verified the reliability of the WD + MS model (\cite{RUI04};
\cite{IHA07}). The purpose of this paper is to study the
progenitor model comprehensively and systematically.

\section{Binary evolution calculation}\label{sect:2}
We use the stellar evolution code of \cite{EGG71, EGG72, EGG73} to
calculate the binary evolutions of WD+MS systems. The code has
been updated with the latest input physics over the last three
decades (\cite{HAN94}; \cite{POL95, POL98}). Roche lobe overflow
(RLOF) is treated within the code described by \cite{HAN00}. Ten
metallicities are chosen here (i.e. $Z=$0.0001, 0.0003, 0.001,
0.004, 0.01, 0.02, 0.03, 0.04, 0.05 and 0.06). The opacity tables
for these metallicties are compiled by \cite{CHE07} from
\cite{IR96} and \cite{AF94}.

Instead of solving stellar structure equations of a WD, we adopt
the prescription in \cite{HAC99a} on the accretion of the
hydrogen-rich material from its companion onto the WD based on an
assumption of optically thick wind (\cite{HAC96}) (see
\cite{MENGXC08} for the prescription in details).

We calculated more than 25,000 WD+MS binary systems with various
metallicities, and obtained a large, dense model grid. We
summarize the final outcomes of all the binary evolution
calculations in the initial orbital period-secondary mass ($\log
P^{\rm i}, M_{\rm 2}^{\rm i}$) planes (see Figs 1 to 10 in
\cite{MENGXC08}). The contours for SNe Ia are shown by solid lines
in these figures. From these figures, we see that the contour
moves from the left lower region to the right upper region with
metallicity Z in the ($\log P^{\rm i}, M_{\rm 2}^{\rm i}$) plane.
This means that the progenitor systems for SNe Ia have a more
massive companion and a longer orbital period for a high Z. We
wrote our results into a FORTRAN code and the code can be
downloaded on X. Meng's personal web site
\emph{http://www.ynao.ac.cn/$^{\rm \sim}$bps/download
/xiangcunmeng.htm}.

\begin{figure}
\begin{minipage}{0.48\textwidth}
\includegraphics[width=50mm,height=65mm,angle=270.0]{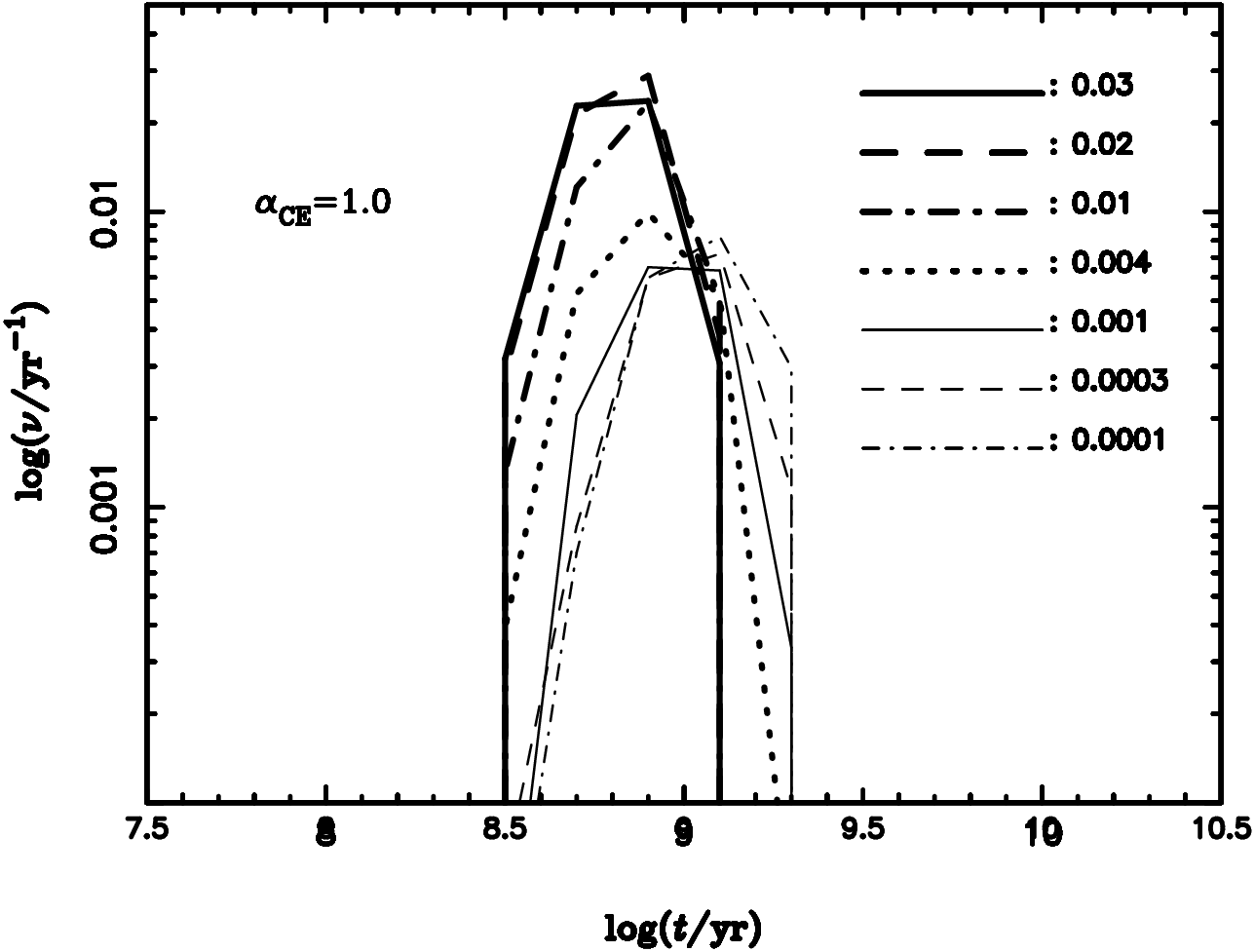}
  \caption{The evolution of the birth rate of SNe Ia for a single
  starburst of $10^{\rm 11}M_{\odot}$ for different metallicities
 with $\alpha_{\rm CE}=1.0$.}\label{fig:birth}
\end{minipage}
\hspace{0.04\textwidth}
\begin{minipage}{0.48\textwidth}
\includegraphics[width=50mm,height=65mm,angle=270.0]{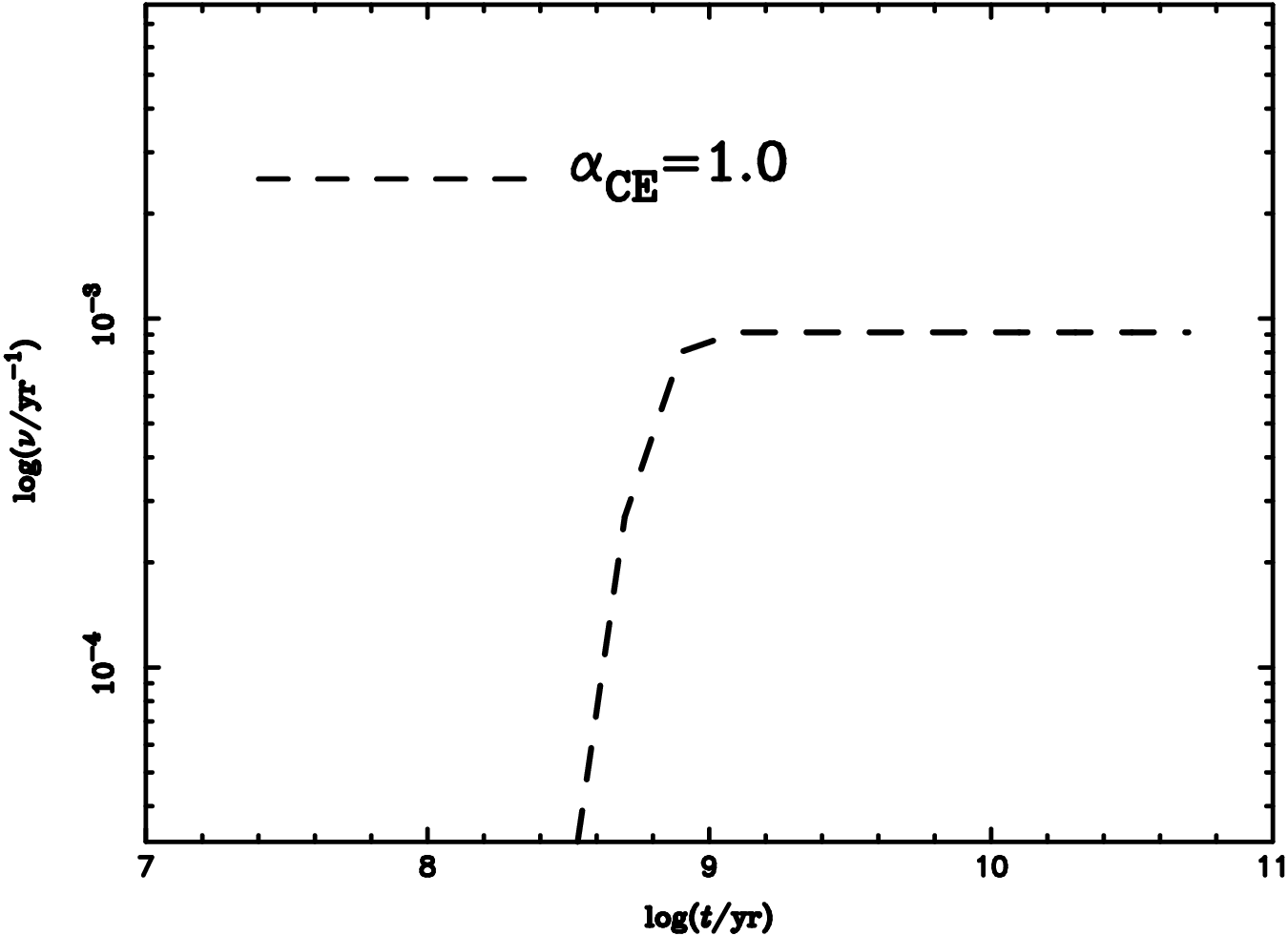}
  \caption{The evolution of the birth rate of SNe Ia for a constant
   star formation rate (Z=0.02, SFR=$5 M_{\rm \odot}{\rm yr^{\rm
-1}}$) with $\alpha_{\rm CE}=1.0$. }\label{fig:sfrbirth}
\end{minipage}
\end{figure}

\section{The results of binary population synthesis}\label{sect:3}
To investigate the birth rate of SNe Ia, we followed the evolution
of $10^{\rm 7}$ binaries for various $Z$ with Hurley's rapid
binary evolution code (\cite{HUR00, HUR02}). The results of grid
calculations in section \ref{sect:2} are incorporated into the
code. Since the code is valid just for $Z\leq0.03$, only seven
metallicities (i.e. $Z=0.03$, $0.02$, $0.01$, $0.004$, $0.001$,
$0.0003$ and $0.0001$) are examined here. The primordial binary
samples are generated in a Monte Carlo way and a circular orbit is
assumed (see \cite{MENGXC08} for details). The results are shown
in Figs. \ref{fig:birth} and \ref{fig:sfrbirth}. From Fig.
\ref{fig:birth}, we see that most supernovae occur between 0.2 Gyr
and 2 Gyr after star formation, and a high metallicity leads to a
systematically earlier explosion time. The peak value of birth
rate increases with metallicity $Z$. However, Fig
\ref{fig:sfrbirth} shows that the WD + MS channel can only account
for about $1/3$ of the Galactic SNe Ia observed(3-4$\times10^{\rm
-3}{\rm yr^{\rm -1}}$, \cite{VAN91}; \cite{CT97}). Therefore,
there may be other channels or mechanisms contributing to SNe Ia.

\begin{figure}
\begin{minipage}{0.48\textwidth}
\includegraphics[width=50mm,height=65mm,angle=270.0]{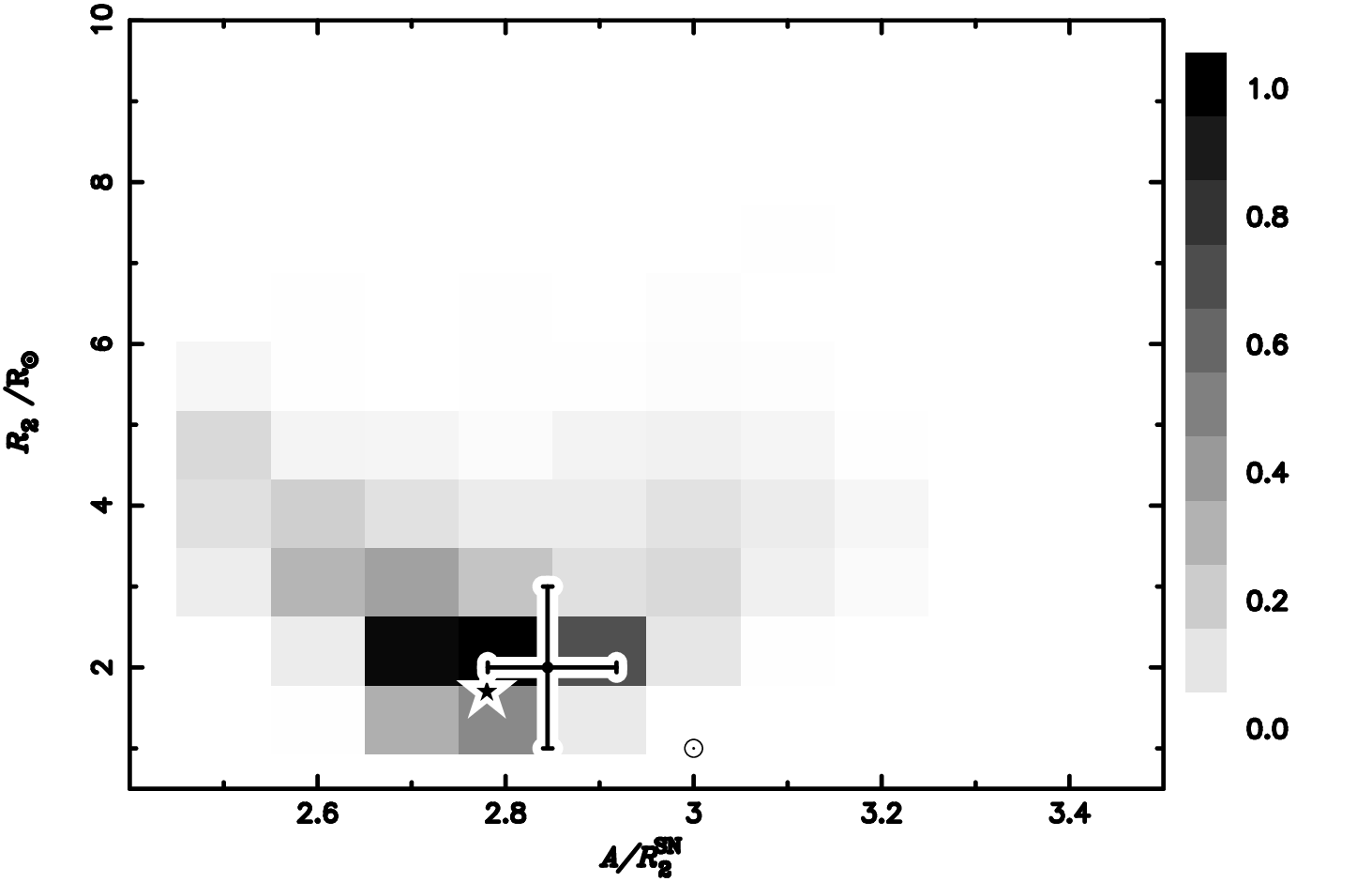}
  \caption{The distribution of the radii and the ratios of
separations to radii of companions at the moment of supernova
explosion for the case of $Z=0.02$ and $\alpha_{\rm CE}=1.0$.
Cross represents the potential candidate of the companion of
Tycho's supernova, Tycho G (\cite{RUI04}; \cite{BRA04}). The
length of the bars of the cross represents observational error.
Solar symbol and pentacle represent the main-sequence and subgiant
companion model in \cite{MAR00}, respectively.
}\label{fig:ratrsn0210}
\end{minipage}
\hspace{0.04\textwidth}
\begin{minipage}{0.48\textwidth}
\includegraphics[width=50mm,height=65mm,angle=270.0]{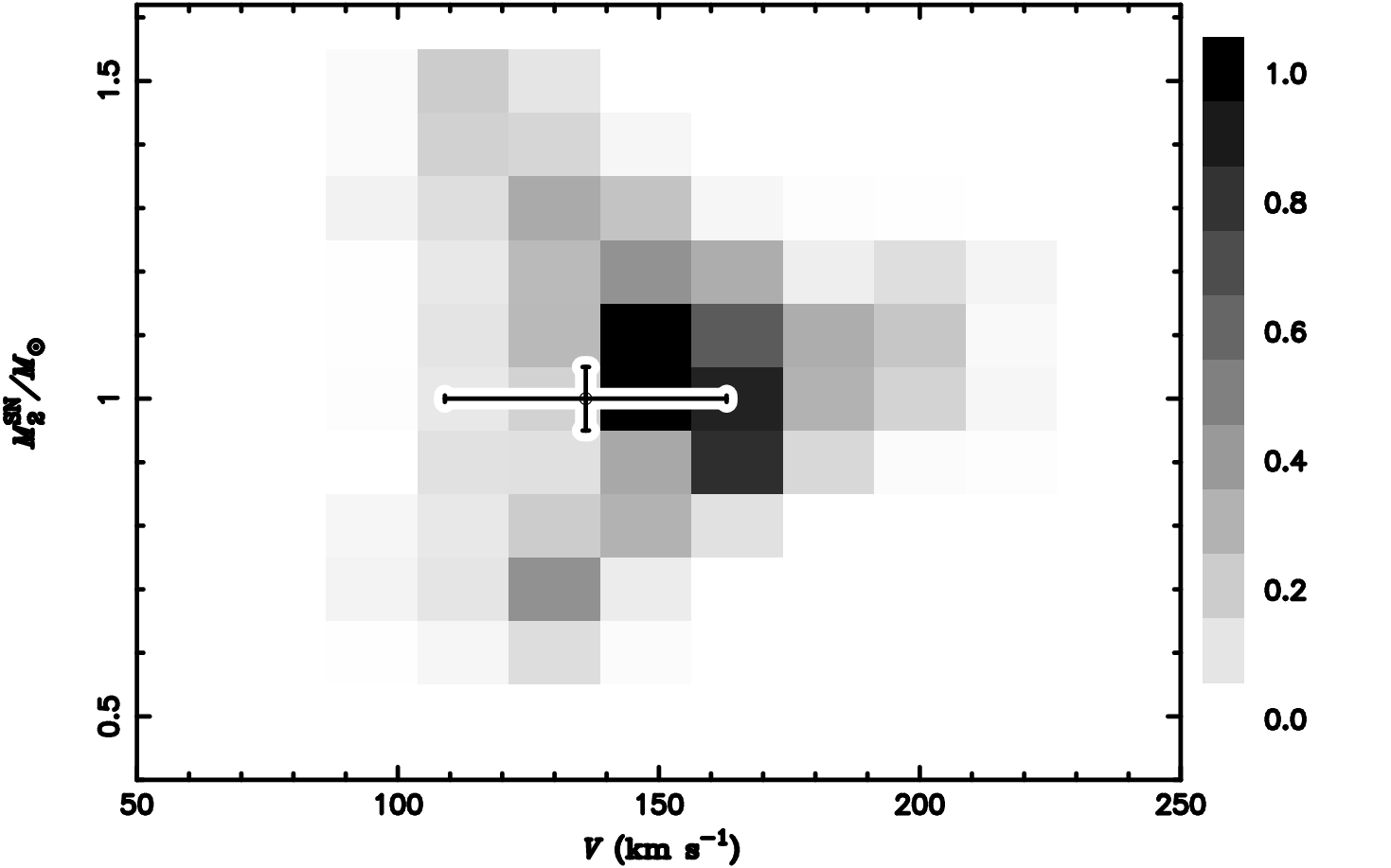}
  \caption{The distribution of the
masses and the space velocities of companions in SN Ia remnants
for $Z=0.02$ and $\alpha_{\rm CE}=1.0$. The cross represents the
potential candidate of the companion of Tycho's supernova, Tycho G
(\cite{RUI04}; \cite{BRA04}), and the length of the bars of the
cross represents observational errors. }\label{fig:vorbmms0210}
\end{minipage}
\end{figure}

We also showed the distributions of the parameters of the binary
systems at the moment of supernova explosion (Fig.
\ref{fig:ratrsn0210}) and the properties of companions after
supernova explosion  (Fig. \ref{fig:vorbmms0210}). The former may
provide physics input when one simulates the interaction between
supernova ejecta and its companion, and the latter may help to
search for the companions in supernova remnants. From Fig.
\ref{fig:vorbmms0210}, we see that the properties of the potential
companion of Tycho's supernova, Tycho G, are well consistent with
our binary population synthesis results. We also noticed from Fig.
\ref{fig:ratrsn0210} that the solar model used by \cite{MAR00}
when they simulated the interaction between explosion ejecta and
the companion is not a typical one (see the solar symbol in the
figure).

A SN Ia remnant is not spherically symmetric due to the companion,
and then the spectrum of the SN Ia may be polarized (\cite{LF05}).
Based on the numerical simulation of \cite{MAR00} and
\cite{KAS04}, we found that at least 75\% of all SNe Ia may be
detected by spectropolarimetry. At present, almost all SNe Ia,
which are observed by spectropolarimetry, have various degrees of
polarization signal (\cite{LEO05}).

Since there exists mass-loss from a progenitor binary system by an
optically thick wind, the material lost from the system may become
circumstellar materials (CSM). The CSM may be the origin of the
color excess of SNe Ia. Based on the single degenerate model with
optically thick wind and using a simple analytic method, we
reproduced the distribution of the color excess of SNe Ia obtained
from observation if a wind velocity of $10$ km/s is adopted. If
the wind velocity is larger than 100 km/s, the reproduction is
bad.

\section{Conclusion}\label{sec:4}
We systematically studied the single degenerate channel of SNe Ia,
(i.e. WD+MS channel), and showed the parameter space leading to
SNe Ia from this channel. Our work may provide help for the future
studies of SNe Ia. We also provided the properties of companions
after supernova explosion, which may help to search for the
companion in SN Ia remnants. We found that at least 75\% of all
SNe Ia may be detected by spectropolarimetry, and then in the
future, statistical studies about the spectropolarimetric
observations of SNe Ia may verify the reliability of the WD + MS
channel.

\begin{acknowledgments}
This work was in part supported by Natural Science Foundation of
China under Grant Nos. 10433030, 10521001, 2007CB815406 and
10603013.
\end{acknowledgments}

\end{document}